# Zero-field quantum anomalous Hall metrology as a step towards a universal quantum standard unit system

Martin Goetz[1], Kajetan M. Fijalkowski[2,3], Eckart Pesel[1], Matthias Hartl[2,3], Steffen Schreyeck[2,3], Martin Winnerlein[2,3], Stefan Grauer[2,3], Hansjoerg Scherer[1], Karl Brunner[2,3], Charles Gould[2,3], Franz J. Ahlers[1]**, Laurens W. Molenkamp[2,3]

[1]*Physikalisch-Technische Bundesanstalt, Bundesallee 100, D-38116 Braunschweig, Germany*
[2]*Physikalisches Institut (EP3), Universität Würzburg, Am Hubland, D-97074, Würzburg, Germany*
[3]*Institute for Topological Insulators, Am Hubland, D-97074, Würzburg, Germany*

**In the quantum anomalous Hall effect, the edge states of a ferromagnetically doped topological insulator exhibit quantized Hall resistance and dissipationless transport at zero magnetic field. Up to now, however, the resistance was experimentally assessed with standard transport measurement techniques which are difficult to trace to the von-Klitzing constant $R_K$ with high precision. Here, we present a metrologically comprehensive measurement, including a full uncertainty budget, of the resistance quantization of V-doped $(Bi,Sb)_2Te_3$ devices without external magnetic field. We established as a new upper limit for a potential deviation of the quantized anomalous Hall resistance from $R_K$ a value of 0.26 ± 0.22 ppm, the smallest and most precise value reported to date. This provides another major step towards realization of the zero-field quantum resistance standard which in combination with Josephson effect will provide the universal quantum units standard in the future.**

Quantum standards are the backbone of the system of measurement units. Already since 1990 all electrical units are based on flux quantization in units of $h/2e$, realized with the Josephson effect [1,2], and conductance quantization in units of $e^2/h$, realized with the quantized Hall effect (QHE) [3,4]. With the revision of the international system of units, SI, in near future [5,6] also the realizations of the units of mass [7,8], the kilogram, and of temperature [9,10], the Kelvin, will utilize and rely on practical electric quantum standards, realizing the vision of Maxwell [11] and Planck [12] of a truly universal system of units. Both electrical quantum standards require temperatures of 4 K or lower for their operation, but since in addition the QHE only works in a magnetic field, it is practically impossible to combine both in one system. However, in ferromagnetic topological insulators like e.g. Cr- or V-doped $(Bi,Sb)_2Te_3$, the quantum anomalous Hall effect (QAHE) provides conductance quantization without a magnetic field [13-16], giving legitimate hope for a future quantum standard where all units based on $h$ and $e$ can be realized in one measurement setup.

Yet, up to now the precision of the QAHE has not been tested with precision metrology methods, and in particular no uncertainty budgets were presented with the data published [17,18]. Indeed, the fact that very low measurement currents are required makes it difficult to reach uncertainties in the parts in $10^9$ range as are routinely obtained in calibrations based on GaAs or graphene QHE devices. A main reason for the limitation of current is the robustness of the ferromagnetic state, which at this stage of development still requires temperatures in the mK-regime and does not tolerate current levels which increase the electron temperature. In the following we present a metrologically comprehensive measurement, including a full uncertainty budget, of the resistance quantization of V-doped (Bi,Sb)$_2$Te$_3$ devices. The device resistance was measured in terms of the von-Klitzing constant derived from a GaAs/AlGaAs heterostructure, using a 100 Ω traveling reference resistor as a transfer device. Both the devices and the reference resistor are described under **Methods**.

---

* Corresponding author: franz.ahlers@ptb.de

**Results**

The measurements were made with a cryogenic current comparator (CCC) bridge very similar to the one which had previously been used for a precision measurement of the Hall quantization in the fractional QHE state [19,20,21]. In that work, the lowest current level of 80 nA had allowed a measurement uncertainty of 6 parts in $10^8$. In the present work on QAHE devices we further decreased the current to as low as 1 nA. Still, an uncertainty of 2.2 parts in $10^7$ was obtained, even with the type-B uncertainty of the reference resistor included, which had been absent in the direct FQHE-to-IQHE comparison. Note that in practical calibration work current levels are typically of the order of 40 µA.

The CCC bridge is schematically shown in Figure 1. One arm of the bridge comprised the QAHE device in a dilution refrigerator (Oxford Triton), connected to a room temperature switching panel by carefully RF-shielded (Q-Filter) leads. A bath temperature of 20 mK was typically used, but depending on current level the electron temperature of the device can be higher. The 100 Ohm reference resistor in the other arm was kept in a regulated thermostat chamber. It is described in more detail in the Methods section. At the core of a feedback loop a DC SQUID detects the flux balance of the coils $N_1$ and $N_2$, with oppositely flowing currents $I_1$ and $I_2$. An auxiliary winding $N_A$ is supplied with a well determined fraction $kI_1$ of the current for compensating the deviation from a perfect integer ratio and balancing the bridge. We used $N_1$=4130, $N_2$=16 and $N_A$=1 in our measurements. From the ratio of the bridge voltage $\Delta U$ to the voltage drop $\Delta(IR)$ across the resistors one obtains the unknown resistance as

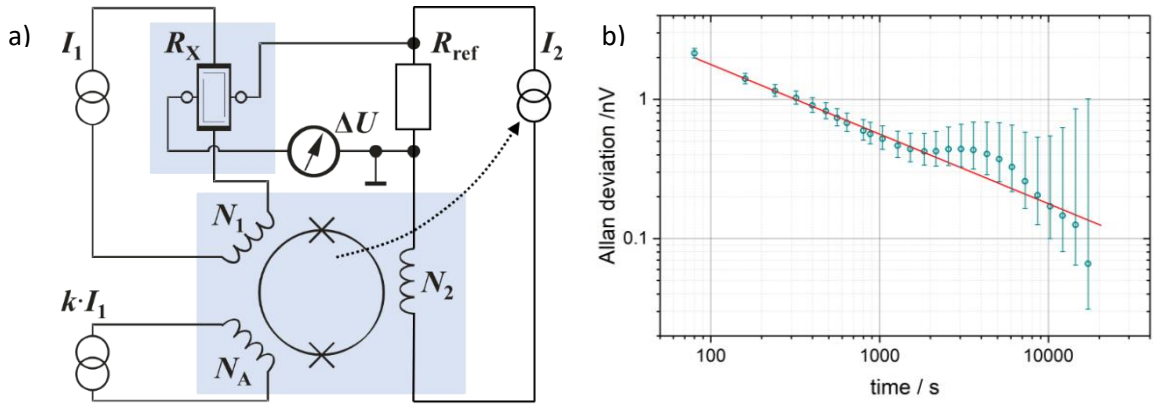

**Figure 1 (a)** Schematics of a CCC bridge. An unknown resistance $R_X$ is measured in terms of a reference resistance $R_{ref}$ by supplying two currents $I_1$ and $I_2$ such that the voltages across $R_{ref}$ and $R_X$ are practically identical. This is achieved by a SQUID-driven feedback loop (symbolized by the dotted array) which adjusts $I_2$ such that the flux through the windings $N_1$, $N_2$, and $N_A$ sensed by the SQUID pick-up loop is zero. A balance of the bridge beyond the finite ratios allowed by $N_1$ and $N_2$ is obtained by injecting a current $kI_1$ into an auxiliary winding $N_A$. **(b)** The Allan deviation of the bridge unbalance voltages of a 5-hour measurement at ±5 nA current follow, within their 95% error bars, a $t^{-1/2}$ law (red line). Already after 15 minutes an uncertainty of 0.5 nV (corresponding to 50 mΩ) is reached. Most of the data in Figure 2 below are based on 15-minute runs, but for the ±5 nA and ±10 nA data several such runs were averaged.

$$R_X = R_{ref} \frac{N_1 + kN_A}{N_2} \left(1 + \frac{\Delta U}{\Delta(IR)}\right) K_{iso}$$

The factor $K_{iso}$ describes the influence of imperfect isolation between connecting leads, or between the control gate and the QAHE device. A deviation from $K_{iso} = 1$ is normally accounted for as a type-B uncertainty contribution derived by determining lower bounds of the relevant isolation resistance. The influence of thermal voltages and their drifts, as well as possible leakage currents to an imperfectly isolated gate, is practically eliminated by the current reversals, which had typical reversal periods of ten seconds, corresponding to an effective measurement frequency of 0.05 hertz. Possible transient artefacts due to current reversals were avoided by discarding the first half of the data points of each reversal half-cycle.

**Table I**  Measurement uncertainty budget listing the relevant uncertainties for a measurement of the quantized anomalous Hall resistance for the case of a current level of ±5 nanoampere. For the individual contributions $x_i$ typical standard errors $s(x_i)$ and their probability distribution functions (PDF) are given, together with the sensitivity coefficients $c(x_i) = \partial R_X / \partial x_i$, the uncertainties $u_i = u(x_i) \cdot c(x_i)$, and $u_i^2$. The relative measurement uncertainty for this specific measurement is obtained as 0.35 ppm.

| $x_i$ | $s(x_i)$ | PDF | $u(x_i)$ | $c(x_i)$ | $u_i$ | $u_i^2$ |
|---|---|---|---|---|---|---|
| $\Delta U$ (type-A) | 7.40·10⁻¹¹ V | *normal* | $s(x_i)$ | 1.000·10⁸ A⁻¹ | 7.40 mΩ | 54.8 (mΩ)² |
| $\Delta U$ (SQUID) | 6.62·10⁻¹¹ V | *rectangular* | $s(x_i)/\sqrt{3}$ | 1.000·10⁸ A⁻¹ | 3.82 mΩ | 14.6 (mΩ)² |
| $K_{iso}$ | 0.09·10⁻⁶ | *rectangular* | $s(x_i)/\sqrt{3}$ | 2.581·10⁴ Ω | 1.28 mΩ | 1.64 (mΩ)² |
| $R_{ref}$ | 2.00·10⁻⁵ Ω | *rectangular* | $s(x_i)/\sqrt{3}$ | 2.581·10² | 2.98 mΩ | 8.88 (mΩ)² |
| | | | | $\sqrt{\sum_i u_i^2}$ | | 8.94 mΩ |
| | | **relative measurement uncertainty: 8.94 mΩ/25.812 kΩ =** | | | | **0.35 ppm** |

The measurement uncertainty budget based on the bridge equation is dominated by the uncertainties of the reference resistor, of the isolation resistance described by the factor $K_{iso}$, and of the uncertainty of the bridge voltage difference $\Delta U$. The latter contributes to the budget with a sensitivity coefficient $\partial R_X / \partial \Delta U \propto 1/\Delta(IR)$. Due to the $I^{-1}$ proportionality and the very low current levels of this experiment all other contributions specific to the bridge hardware (i.e. the winding ratio $(N_1 + kN_A)/N_2$) contribute roughly three orders of magnitude less than $\Delta U$ and are therefore not listed in Table I. The $\Delta U$-uncertainty itself comprises contributions from detector noise, SQUID noise, thermal noise of the resistors and contacts, and potentially from noise pick-up of the cabling, all of which have a normal probability distribution function (PDF) and can be determined from the Allan deviation of the measured signal time traces (Figure 1 b) gives an example). The averaged value from several 15-minute measurements at 5 nanoampere is given as the standard deviation in line 1 of the uncertainty budget in Table I. The type-B uncertainty contribution due to the SQUID non-linearity, described in detail in [22], also contributes to the uncertainty of $\Delta U$, but with a rectangular PDF. It is listed separately in line 2. The last two lines give the type-B uncertainties of $K_{iso}$ and $R_{ref}$, both also having a rectangular PDF.

In this table $K_{iso}$ was estimated by determining the resistance of one lead in the cryostat against all others to be $R_{iso}$ = (300 ± 50) GΩ. From the high potential lead to the Hall bar, a fraction of (25.8kΩ/300GΩ = 0.09 ppm) of the current can bypass the Hall bar, leading to an apparent underestimate of the measured resistance by the same fraction. In our final result which we give below we will correct the measured resistance by 0.09 ppm, but still keep the full uncertainty of $K_{iso}$ in the budget. For $R_{ref}$ a relative uncertainty of 0.2 ppm (corresponding to 20 µΩ) was used, which is a very conservative estimate, considering the excellent stability of the temporal drift of that resistor during the weeks before and after the measurements done in Würzburg. This drift is shown in Figure 6 in section Methods.

All precision measurements were performed with the larger of the two Hall bars described in section Methods, where also the selection of the gate voltage is described. The result of these measurements is shown in Figure 2. It summarizes measurements of $R_{xy}$ at zero magnetic field taken with currents of 1, 2, 5, 10, 20 and 50 nA, using the 100 Ω resistor as the reference. At 5 and 10 nA each, several measurements were made (more at 5 nA than at 10 nA) and their weighted average is plotted. At all other currents only one measurement was made. The 5 and 10 nA data comprise data from both orthogonal contact pairs (2,4) and (1,3) in Figure 4, while the other data stem from contact pair (2,4). Note that one reading, marked in green color, was taken at a field of 2 T and deviates significantly from the other data, which were taken after ramping the field to zero and waiting for more than 2 hours to achieve stable readings.

To the data in Figure 2 we fitted a power law of the form $a + b(I/\text{nA})^c$ which had successfully been used in extrapolating the Hall resistance of a GaAs-based QHE device in the fractional QH state at filling factor 1/3 recently [19] and which also has been observed for an integer QHE device [23]. Interestingly, in all cases an exponent of 3 is observed in this power law, for the cited IQHE and FQHE measurements as well as for the QAHE result obtained here. While one might naively expect that a deviation from exact quantization with increasing current should in first order be proportional to $I^2$ due to the dissipated power, one should also keep in mind that an increasing Hall electric field, proportional to $I$, can induce scattering leading to breakdown of quantization. Although this reasoning is far from a theoretical modeling, it might explain the observed exponent of 3. The fact that the QAHE is in this respect very similar to both integer and fractional QHE regimes indicates that the mechanisms for the onset of breakdown are probably similar, despite the different physics of these cases.

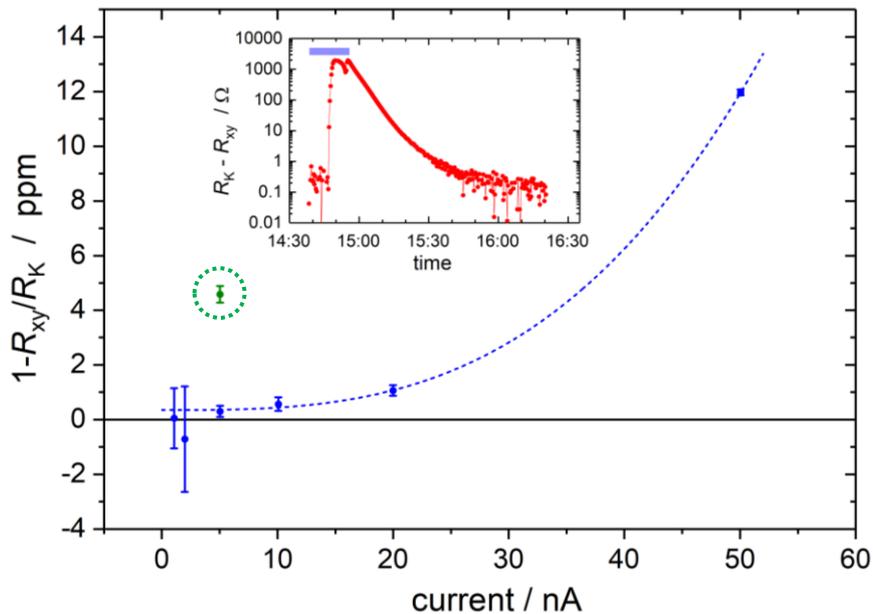

**Figure 2** Measurement of the Hall resistance of the larger of the two Hall bars against a 100 Ω transfer standard in dependence on current. The device was at a temperature of 20 mK at zero magnetic field. The dashed blue curve represents a power law fit of the form $a + b(I/\text{nA})^c$ with $a = (0.35 \pm 0.10)$ ppm and $c = 3.02 \pm 0.20$. The value of 4.5 ppm at 5 nA current highlighted by a green circle was measured with a magnetic field $B$ of 2 T, before the field was ramped to zero where all other measurements were taken. The development of $R_{xy}$ during and after the field ramping is shown in the inset, where the blue bar at the top indicates the period over which $B$ was changed.

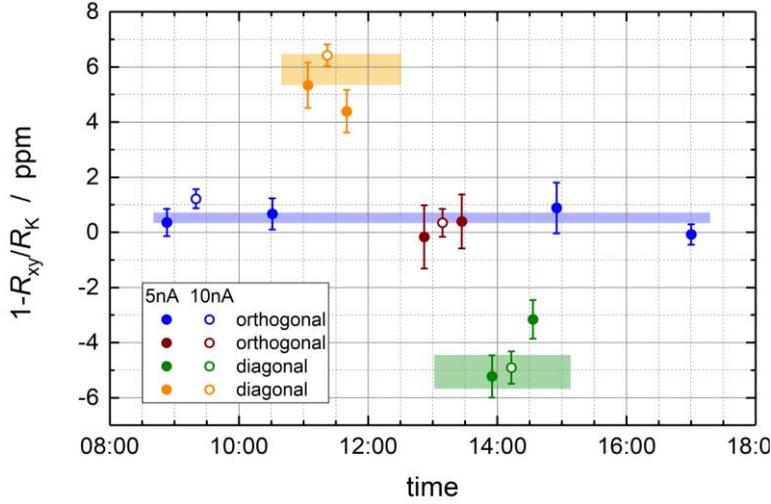

**Figure 3** Series of measurements of $1 - R_{xy}/R_K$. Measurement currents of 5 nA and 10 nA were used in orthogonal (contact pair [1,3] or [2,4] in Figure 4) and diagonal (contact pair [1,4] or [2,3]) contact configurations. The colored rectangles represent the (weighted) average and standard deviation of the orthogonal and diagonal groups of data points.

The central result from Figure 2 is, however, the value $a = (0.35 \pm 0.10)$ ppm for the extrapolated-to-zero Hall resistance. As mentioned above, we correct this value by the influence of the imperfect lead isolation of 0.09 ppm. We also must add to the uncertainty of 0.1 ppm (which reflects only the type-A uncertainties of the measurement) the uncertainty contributions from lines 2 to 4 of Table I. Doing so we obtain a value of

$$1 - \frac{R_{xy}}{R_K} = (0.26 \pm 0.22) \cdot 10^{-6}$$

At currents of 5 and 10 nA also measurements of the longitudinal resistance $R_{xx}$ were made. We used the difference of bridge readings obtained from an orthogonal contact pair (e.g. (2,4)) and a diagonal pair (e.g. (1,4)) as a value for $R_{xx}$. This method provides more reliable results than a direct reading of the longitudinal voltage with the bridge detector, since it does not suffer from the finite input resistance of that instrument. Figure 3 shows results from a total of 14 measurements taken over 8 hours, using different potential contact pairs. The colored bars indicate the average and standard deviation of respective groups of contacts, with a blue rectangle for the orthogonal contacts and orange and green rectangles for the diagonal configurations. From the figure, we can read a longitudinal resistance of 5.5 ppm of $R_K$, equivalent to 142 mΩ, or 47 mΩ☐. We do, however, not find a difference between the 5 nA and 10 nA measurements in this initial assessment.

**Conclusion**

In conclusion, we have performed a metrologically sound precision measurement of the quantized anomalous Hall resistance of the Vanadium doped topological insulator $(Bi,Sb)_2Te_3$. Despite the low measurement currents below 10 nA excellent agreement with the von-Klitzing-constant $R_K$ was found. Since the development of this class of materials is just at the beginning, it can be expected that better ferromagnetic TI material quality with higher Curie temperatures will become available, leading to considerably higher operation temperature and breakdown currents which would allow measurement uncertainties competing with those from GaAs-based QHE devices. The distinctive advantage of a QAHE resistance standard, its operation without a magnetic field, would then make the other quantum resistance standards obsolete, which require magnetic fields for their operation. In addition, the zero-field operation would allow to integrate Josephson voltage and QAHE resistance standards in one setup, thus yielding the ultimate electrical standard which supplies voltage, resistance, and (by virtue of Ohm's law) current from one reference instrument.

## Methods

**Sample fabrication**

The devices we used were made of 9 nm thick films of the ferromagnetic topological insulator $V_{0.1}(Bi_{0.21} Sb_{0.79})_{1.9}Te_3$, grown by molecular beam epitaxy (MBE) on a hydrogen passivated Si(111) substrate and covered in-situ by a 10 nm thick Te protective cap [24]. The devices were patterned using standard optical lithography. To allow for top gating, dielectric layers of 20 nm $AlO_x$ and 1 nm $HfO_x$ were deposited by atomic layer deposition, and covered with 5 nm of Ti and 100 nm of Au. Six-terminal Hall bar devices were patterned by Ar ion beam etching. As shown in Figure 4, a sample contains two Hall bars of aspect ratio 3:1, with widths of 200 μm and 10 μm, respectively. For contact metal deposition, the Te-cap was removed by argon ion beam etching before transferring it to the metallization chamber under continuous high vacuum conditions, where 50 nm AuGe, 5 nm Ti, and 50 nm of Au was deposited. The side contacts were 20 μm wide for the bigger Hall-bar, with typical contact resistances of 300 Ω. The sample was glued into a chip carrier and connected using standard "wedge" bonding. All measurements shown in this paper were made with the bigger Hall bar shown in the right part of Figure 4**.**

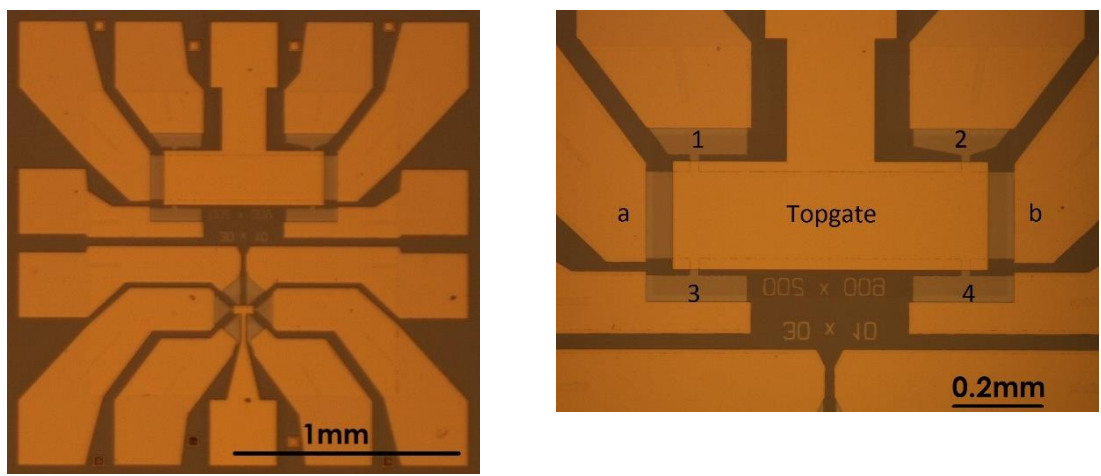

**Figure 4** Micrograph of the sample used for the measurements. The left figure shows an overview with the two Hall bars of different size. The right figure shows an enlarged view of the bigger Hall bar with current contacts labelled a and b, and potential contacts labelled 1 to 4. Note that the potential contacts are much closer to the current contacts than in usual Hall bars employed for precision measurement. They are thus potentially more susceptible to current induced breakdown.

**Gate voltage dependence**

As a first step before the actual precision measurement, and after coarse balancing of the bridge, a gate voltage scan was performed on both Hall bars shown in Figure 4. Under step-wise change of the gate voltage the bridge voltage difference revealed, as expected, a wider and more stable quantized resistance plateau (in dependence on gate voltage) for the larger Hall bar. All measurements shown in the main section were performed with this Hall bar, using a gate voltage of 5.5 V.

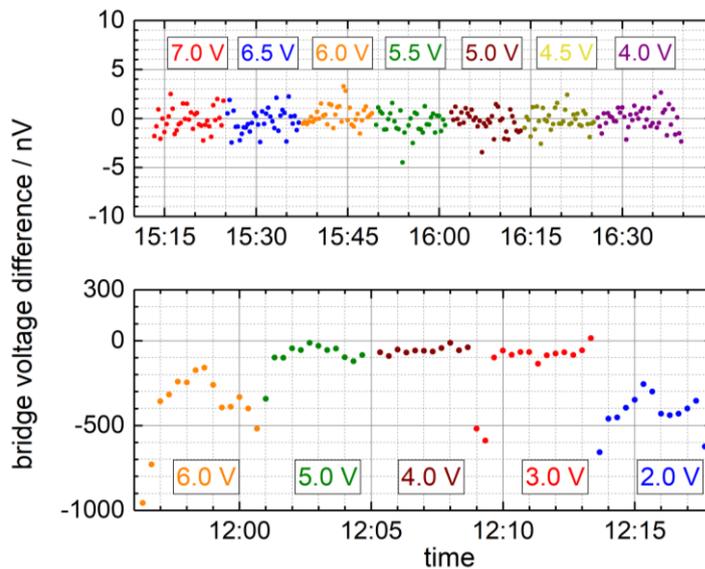

**Figure 5** Bridge voltage differences in dependence on gate voltage for the large Hall bar (upper panel) and the small Hall bar (lower panel). At zero magnetic field and with a measurement current of ±5 nanoampere the voltage difference of the coarsely balanced bridge was continuously monitored while the gate voltage was changed in steps of 0.5 V (upper panel) and 1 V (lower panel). Color coding identifies the readings at a given gate voltage. Note the different scales of the ordinates.

**Reference resistor**

The reference resistor used in our measurements has a nominal value of 100 Ω. During the measurements, it was kept in a regulated thermostat chamber at 23.1 ± 0.05 °C, corresponding to an uncertainty of 0.02 ppm in this control range. Its value is calibrated against the GaAs-based German resistance standard at PTB in regular intervals, which are plotted in Figure 6 around the time of the measurements performed in Würzburg. Despite the excellent stability seen from this diagram we assigned a relative uncertainty of 0.2 ppm to the reference standard during the 3 days measurement campaign in Würzburg, where no calibration against a GaAs reference was performed.

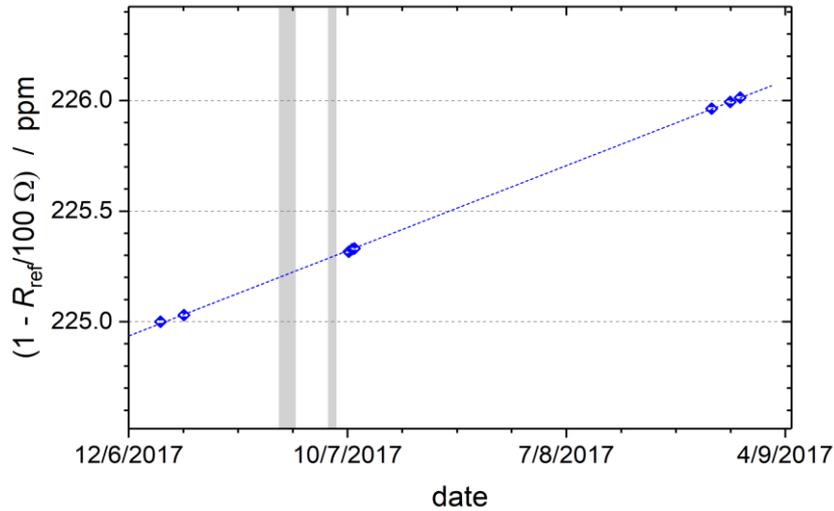

**Figure 6** Drift behavior of the traveling resistance standard around the time of measurement. The grey areas indicate the time when the resistor was moved from Braunschweig to Würzburg and back. During these periods the resistor underwent temperature excursions of up to ±5 K and was also exposed to mechanical shock typical for a regular transport by car. The individual calibration values have uncertainties smaller than the blue diamond symbols.

## Acknowledgements

We gratefully acknowledge the financial support of the EU ERC-AG Program (project 3-TOP), the DFG through SFB 1170 (ToCoTronics), the Leibniz Program, and the Elitenetzwerk Bayern IDK Topologische Isolatoren. MG, EP, HS and FJA would like to thank Ingrid Mertig for stimulating and encouraging discussions.

## Author contributions

S.G. fabricated the sample; M.W. and S.S. grew the layers; M.G., E.P., K.M.F. and M.H. performed the measurements and analysis; K.B., C.G., H.J.S., F.J.A., and L.W.M. conceived of the experiment and planned its execution. All authors discussed the results and participated in the preparation of the manuscript.

## Additional information

Competing interests: The authors declare no competing financial interests.

## References

[1] Josephson, B. D. Possible new effects in superconductive tunnelling. Physics Letters, **1**, 251 (1962).

[2] Shapiro, S. Josephson Currents in Superconducting Tunneling: The Effect of Microwaves and Other Observations. Physical Review Letters, **11**, 80 (1963).

[3] von Klitzing, K., Dorda, G., Pepper, M. New method for high-accuracy determination of the fine-structure constant based on quantized Hall resistance. Phys. Rev. Lett. **45**, 494 (1980).

[4] Delahaye, F. & Jeckelmann, B. Revised technical guidelines for reliable dc measurements of the quantized Hall resistance. Metrologia **40**, 217 (2003).

[5] Mills, I.M., Mohr, P. J., Quinn, T. J., Taylor, B. N. & Williams, E. R. Redefinition of the kilogram, ampere, kelvin and mole: a proposed approach to implementing CIPM recommendation 1 (CI-2005). Metrologia **43**, 227 (2006).

[6] BIPM information 'On the future revision of the SI': www. bipm.org/en/measurement-units/rev-si/

[7] Kibble, B. P. A Measurement of the Gyromagnetic Ratio of the Proton by the Strong Field Method. Atomic Masses and Fundamental Constants, **5**,545 (1976).

[8] Stock, M. Watt balance experiments for the determination of the Planck constant and the redefinition of the kilogram. Metrologia **50**, R1 (2013)

[9] The Consultative Committee for Thermometry, Recommendation T 1 (2014) on a new definition of the kelvin www.bipm.org/cc/CCT/Allowed/Summary_reports/RECOMMENDATION_web_version.pdf (2014)

[10] Qu, J. et al. An improved electronic determination of the Boltzmann constant by Johnson noise thermometry. Metrologia **54**, 549 (2017).

[11] Maxwell, J.C. British Association Report, XL (1870), reproduced in: The Scientific Papers of James Clerk Maxwell **2**, 225 (1890)

[12] Planck, M. Sitzungsberichte der Königlich Preußischen Akademie der Wissenschaften zu Berlin (18 May 1899), **440**, reprinted in: Physikalische Abhandlungen und Vorträge (Vieweg, Braunschweig), **I**, 560 (1958).

[13] Chang, C.-Z. et al. Experimental observation of the quantum anomalous Hall effect in a magnetic topological insulator. Science **340**, 167 (2013).

[14] Checkelsky, J. G. et al. Trajectory of anomalous Hall effect toward the quantized state in a ferromagnetic topological insulator. Nature Physics **10**, 731 (2014).

[15] Bestwick, A. J. et al. Precise quantization of the anomalous Hall effect near zero magnetic field. Phys. Rev. Lett. **114**, 187201 (2015).

[16] Grauer, S. et al. Coincidence of superparamagnetism and perfect quantization in the quantum anomalous Hall state. Phys. Rev. B **92**, 201304 (2015).

[17] Chang, C.-Z. et al. High-precision realization of robust quantum anomalous Hall state in a hard ferromagnetic topological insulator. Nat. Mater. **14**, 473 (2015).

[18] During writing of this manuscript, we became aware of the preprint Fox, E. J. et al. Part-per-million quantization and current-induced breakdown of the quantum anomalous Hall effect. arXiv:1710.01850 which presents similar results on Cr-doped (Bi,Sb)2Te3 samples.

[19] Ahlers, F. J., Götz, M. & Pierz, K. Direct comparison of fractional and integer quantized Hall resistance. Metrologia **54**, 516 (2017).

[20] Drung, D. et al. Improving the stability of cryogenic current comparator setups. Supercond. Sci. Technol. **22**, 114004 (2009).

[21] Drung D. & Storm, J.-H. Ultralow-noise chopper amplifier with low input charge injection. IEEE Trans. Instrum. Meas. **60**, 2347 (2011).

[22] Drung, D., Götz, M., Pesel, E. & Scherer, H. Improving the traceable measurement and generation of small direct currents- IEEE Trans. Instrum. Meas. **64**, 3021 (2015).

[23] Ribeiro-Palau, R. et al. Quantum Hall resistance standard in graphene devices under relaxed experimental conditions (Supplementary figure 3). Nat. Nanotechnol. **10**, 965 (2015).

[24] Winnerlein, M. et al. Epitaxy and structural properties of (V,Bi,Sb)$_2$Te$_3$ layers exhibiting the quantum anomalous Hall effect. Phys. Rev. Materials **1**, 011201(R) (2017).